%% file: poster_maizapellanizj2.tex
\documentclass[twoside,a4paper,11pt]{sea10}
\usepackage{graphicx}
\usepackage{hyperref}
\usepackage{movie15}
\input{mlaeff}

\begin{document}
\pagenumbering{arabic}
\pagestyle{myheadings}
\thispagestyle{empty}
{\flushleft\includegraphics[width=\textwidth,bb=58 650 590 680]{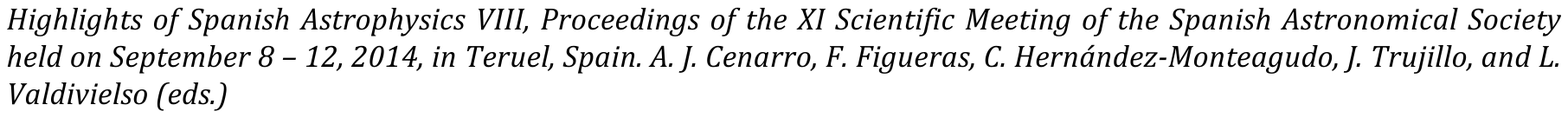}}
\vspace*{0.2cm}
\begin{flushleft}
{\bf {\LARGE
%
The Hourglass as seen with HST/WFPC2
%
}\\
\vspace*{1cm}
%
J. Ma{\'\i}z Apell{\'a}niz$^1$, 
L. {\'U}beda$^2$,
R. H. Barb{\'a}$^3$,
J. W. MacKenty$^2$,
J. I. Arias$^3$,
and 
A. I. G{\'o}mez de Castro$^4$
%
}\\
\vspace*{0.5cm}
%
$^1$
Centro de Astrobiolog{\'\i}a, INTA-CSIC, Spain\\
$^2$
Space Telescope Science Institute, USA\\
$^3$
Universidad de La Serena, Chile\\
$^4$
Universidad Complutense de Madrid, Spain
%
\end{flushleft}
%
\markboth{
The Hourglass as seen with HST/WFPC2
}{ 
%
Ma{\'\i}z Apell{\'a}niz et al.
%
}
\thispagestyle{empty}
\vspace*{0.4cm}
\begin{minipage}[l]{0.09\textwidth}
\ 
\end{minipage}
\begin{minipage}[r]{0.9\textwidth}
\vspace{1cm}
\section*{Abstract}{\small
%
We present a multi-filter HST/WFPC2 UV-optical study of the Hourglass region in M8. We have extracted the stellar photometry of
the sources in the area and obtained the separations and position angles of the Herschel 36 multiple system: for Herschel 36 D we
detect a possible orbital motion between 1995 and 2009. We have combined our
data with archival IUE spectroscopy and measured the Herschel 36 extinction law, obtaining a different result from that of
\href{http://adsabs.harvard.edu/abs/1989ApJ...345..245C}{Cardelli et al. (1989)} due to the improvement in the quality of the 
optical-NIR data, in agreement with the results of 
\href{http://adsabs.harvard.edu/abs/2014A&A...564A..63M}{Ma{\'\i}z Apell{\'a}niz et al. (2014)}.  A large fraction of the UV flux 
around Herschel 36 arises from the Hourglass and not directly from the star itself. In the UV the Hourglass appears to 
act as a reflection nebula located behind Herschel 36 along the line of sight. Finally, we also detect three new Herbig-Haro objects
and the possible anisotropic expansion of the Hourglass Nebula.
%
\normalsize}
\end{minipage}
\begin{figure}
\centerline{\includegraphics*[width=1.00\linewidth, bb=0 0 533 533]{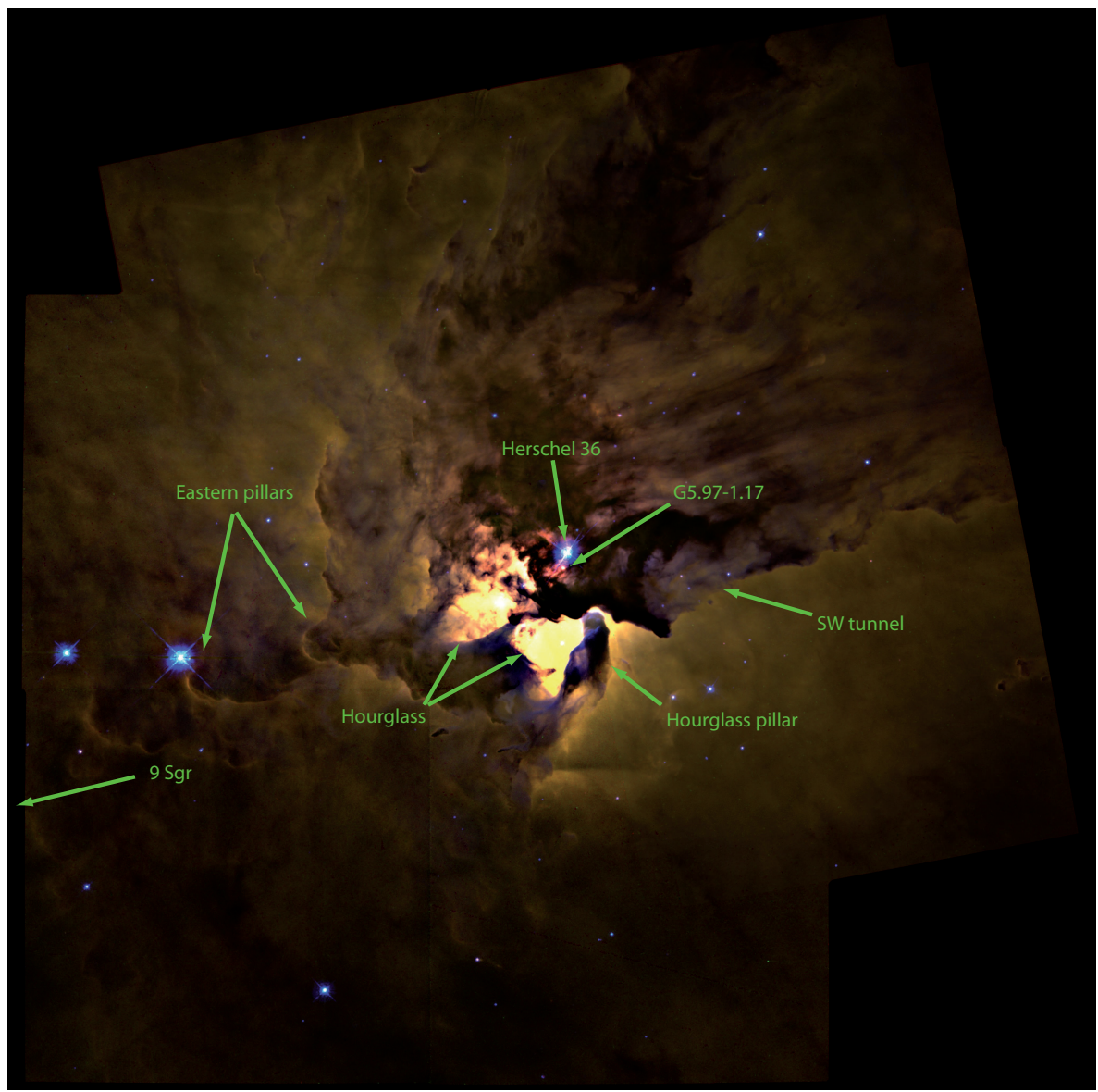}}
\caption{F656N (red) + F487N (green) + F547M (blue) WFPC2 mosaic of the extended region around the Hourglass. Redder regions indicate higher
extinction in the nebular lines. In bluish regions scattered radiation is more significant. The field size is 205\arcsec~$\times$~205\arcsec\
(1.3~pc~$\times$~1.3~pc) and N is 45\degr\ CCW from top.}
\label{fig1}
\end{figure}

\section{Motivation}

$\,\!$\indent Herschel 36 (O7: V + sec, \href{http://adsabs.harvard.edu/abs/2014ApJS..211...10S}{Sota et al. 2014}) is a unique object.
It is the main ionizing source of the Hourglass Nebula (for most of M8 it is 9 Sgr),
a ZAMS SB3 system (O7.5 V + O9 V + B0.5 V, \href{http://adsabs.harvard.edu/abs/2010ApJ...710L..30A}{Arias et al. 2010}), and
the prototype of large-$R_{5495}$ extinction laws (\href{http://adsabs.harvard.edu/abs/1989ApJ...345..245C}{Cardelli et al. 1989}).
Herschel 36 C (0\farcs25 away) is another B star companion, deeply embedded in dust 
         (\href{http://adsabs.harvard.edu/abs/2006ApJ...649..299G}{Goto et al 2006} and contribution by J. Ma{\'\i}z Apell\'aniz in these proceedings).
In the surrounding nebula,
the Hourglass region has a very high surface brightness,
the gas and dust distributions are complex (\href{http://adsabs.harvard.edu/abs/2006MNRAS.366..739A}{Arias et al. 2006}), and 
several other interesting objects (\href{http://adsabs.harvard.edu/abs/2006ApJ...649..299G}{Goto et al. 2006}, 
         \href{http://adsabs.harvard.edu/abs/2006MNRAS.366..739A}{Arias et al. 2006}) can be found.
All of this prompted us to request HST time during Supplemental Cycle 16, which was the last chance for FUV observations with WFPC2.

\section{The data}

$\,\!$\indent We used three kinds of data:

\begin{itemize}
 \item Archival WFPC2 data (GO 6227, 1995, PI: Trauger). 
       These have the Hourglass in the PC, with WF2 to the W and 
       include $V$ + $I$ filters (F547M + F814W) and
       nebular filters (F487N + F502N + F656N + F658N + F953N).
 \item New WFPC2 data (GO 11\,981, 2009, PI: Ma{\'\i}z Apell\'aniz).
       These have the Hourglass in the PC, with WF2 to the E and
       include new FUV to $B$-band filters (F170W + F255W + F336W + F439W),
       an $R$-band filter (F675W), and 
       previously used filters for larger coverage and second epochs (F487N + F547M + F656N + F814W).
 \item Archival 2MASS $JHK_s$ photometry and IUE spectroscopy for Herschel 36.
\end{itemize}

\begin{figure}
\centerline{\includegraphics*[width=1.00\linewidth, bb=0 0 162 159]{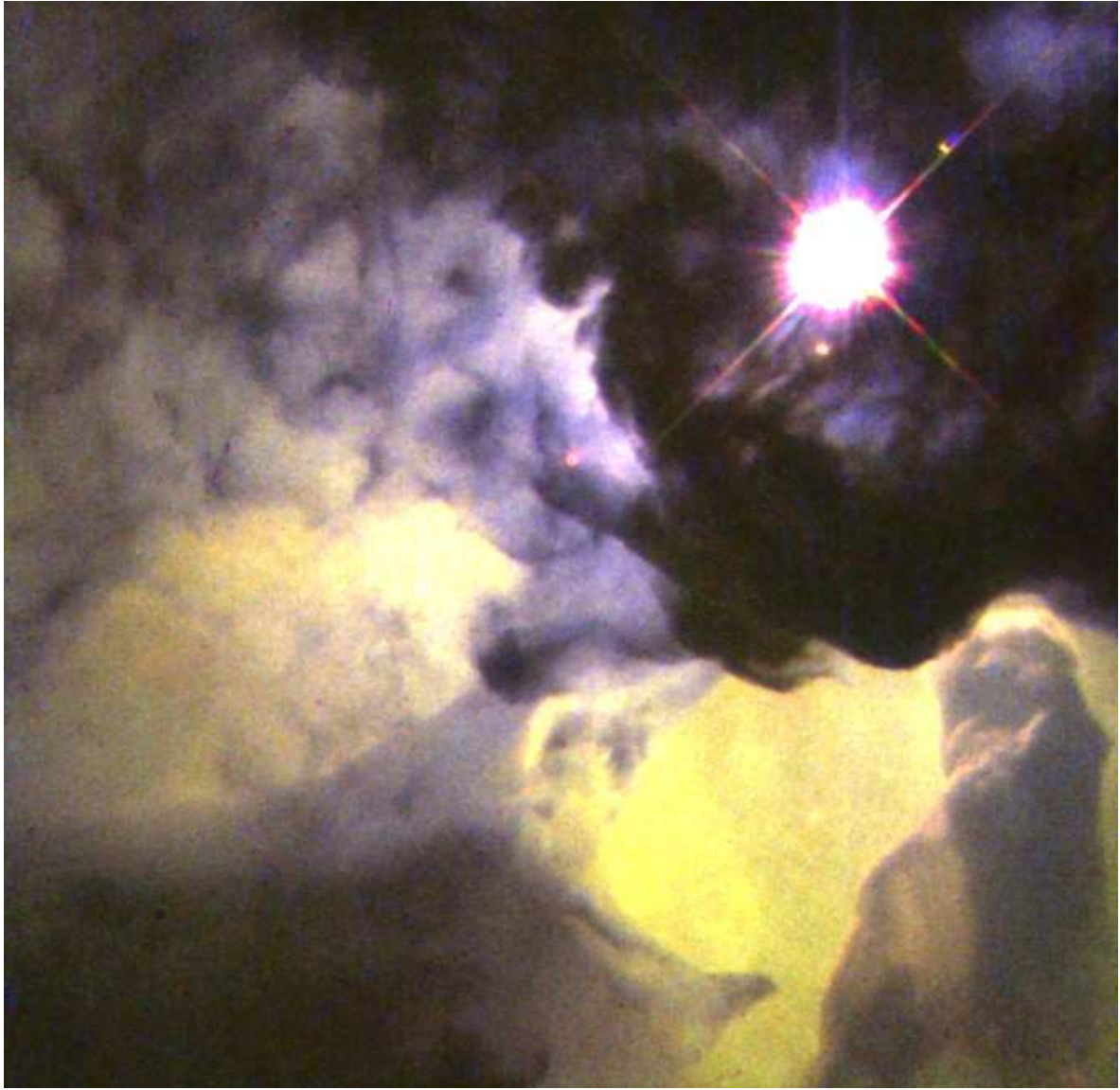}}
\caption{F439W (red) + F336W (green) + F170W (blue) PC mosaic of the Hourglass. Herschel 36 is the bright point source in the upper right
quadrant. The extended F439W and F336W is mostly of nebular origin while F170W is mostly reflected light. The field size is 
34\arcsec~$\times$~34\arcsec\ (0.22~pc~$\times$~0.22~pc) and N is 45\degr\ CCW from top.}
\label{fig2}
\end{figure}

\section{Processing}

$\,\!$\indent See \href{http://adsabs.harvard.edu/abs/2006MNRAS.366..739A}{Arias et al. (2006)} for the initial processing of GO 6227 data.
We performed aperture photometry for point-like or quasi-point-like sources
using the original (geometrically-distorted) data
and selecting the best exposure time as a function of magnitude.
For saturated sources we applied techniques similar to that of \href{http://adsabs.harvard.edu/abs/1994ApJ...435L..63G}{Gilliland (1994)}
         for GAIN=15 and \href{http://adsabs.harvard.edu/abs/2003hstc.conf..346M}{Ma{\'\i}z Apell\'aniz (2003)} for GAIN=7. 
We used a realistic (spatially-varying) background subtraction and
applied CTI, contamination, and aperture corrections.
Aperture and zero-point uncertainties were added to the final result. 
A ghost produced by Herschel 36 was discarded in the long exposures.
Finaly, the nebulosity was analyzed with large-area photometry.

\begin{figure}
\centerline{\includegraphics*[width=1.00\linewidth, bb=100 430 591 824]{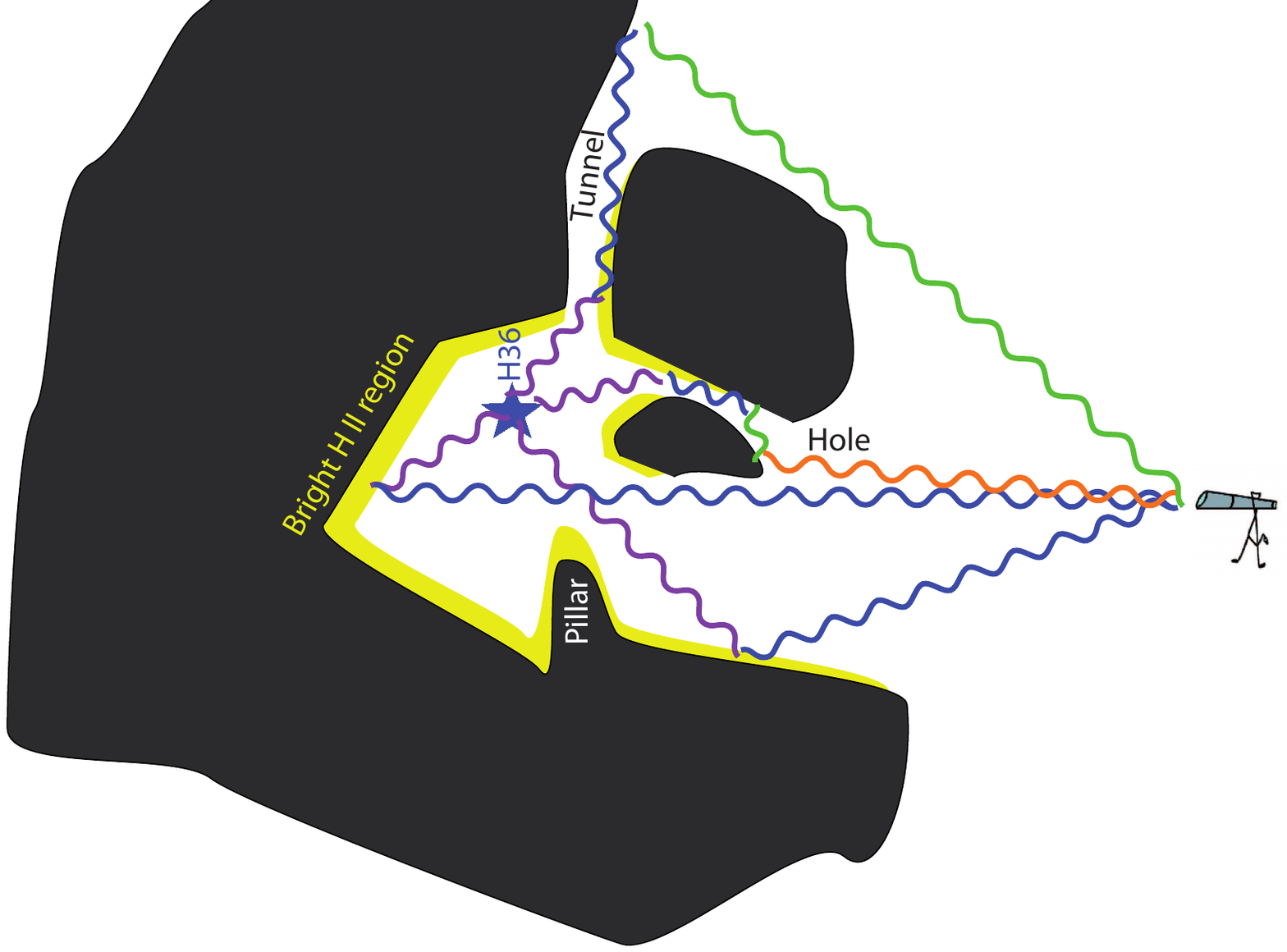}}
\caption{A toy model of the Hourglass region in M8 seen in cross section with respect to the plane of the sky. The path followed by 
radiation is represented by progressively redder lines as it is scattered in different parts of the nebula. The telescope position marks the 
direction of our point of view. The effect of external ionizing sources such as 9~Sgr is not included.}
\label{fig3}
\end{figure}

\section{The overall structure of M8}

$\,\!$\indent As seen in Figure~\ref{fig1}, the Hourglass has a much higher surface brightness than the rest of M8.
The eastern pillars point towards 9 Sgr (O4 V((f))z, \href{http://adsabs.harvard.edu/abs/2014ApJS..211...10S}{Sota et al. 2014}), 
       the main ionizing source of M8.
The southern limit of the Hourglass is a pillar pointing towards Herschel 36. 
The region around Herschel 36 shows higher extinctions than the Hourglass (Figure~\ref{fig1} and 
       \href{http://adsabs.harvard.edu/abs/2006MNRAS.366..739A}{Arias et al. 2006)}: reddened holes are seen among 
       (even more extinguished) dark regions.
Some regions (tunnel towards the SW, part of the Hourglass pillar) are relatively brighter in F547M with respect to F656N or F487N: they are 
       illuminated by mostly non-ionizing radiation. 

\begin{table}
\caption{Results of the CHORIZOS fits for Herschel 36.}
\centerline{$\,\!$}
\centerline{
\begin{tabular}{lcc}
\hline
Quantity                     & CCM laws        & New laws        \\
\hline
$\chi^2_{\rm red}$           & 5.3             & 2.0             \\
$E(4405-5495)$               & 0.883$\pm$0.008 & 0.784$\pm$0.008 \\
$R_{5495}$                   & 5.098$\pm$0.073 & 5.942$\pm$0.096 \\
$\log d^*$                   & 3.049$\pm$0.007 & 3.023$\pm$0.007 \\
$E({\rm F439W}-{\rm F547M})$ & 0.954$\pm$0.008 & 0.835$\pm$0.008 \\
$A_{\rm F547M}$              & 4.512$\pm$0.037 & 4.670$\pm$0.038 \\
${\rm F547M}_0$              & 5.743$\pm$0.033 & 5.613$\pm$0.035 \\
\hline
\end{tabular}
}
$^*$ Do not trust: it assumes a single, typical MS star.
\bigskip
\label{tab1}
\end{table}

\begin{figure}
\centerline{\includegraphics*[width=0.80\linewidth, bb=28 28 566 566]{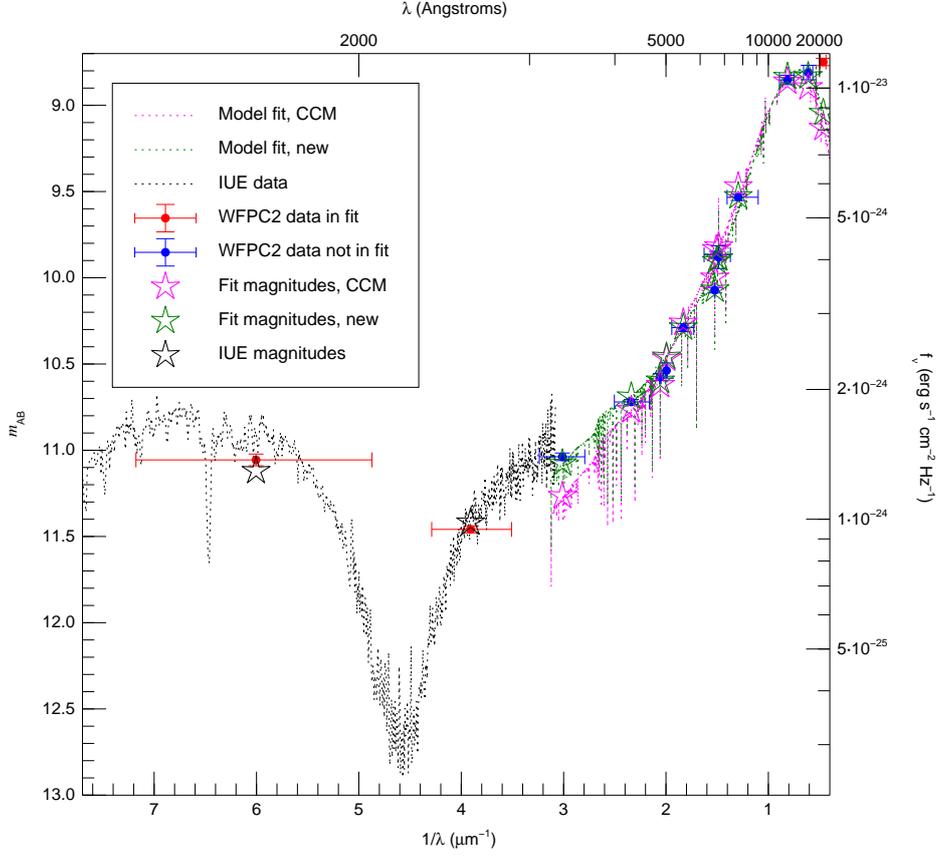}}
\caption{WFPC2+2MASS photometry, CHORIZOS fits to the optical+NIR photometry using the CCM and new families of extinction laws, and IUE
spectroscopy of Herschel 36. The synthetic magnitudes are also shown for the two fits and the IUE spectroscopy. Note that the new extinction laws
provide a better fit than the CCM ones.}
\label{fig4}
\end{figure}

\begin{figure}
\centerline{\includegraphics*[width=0.80\linewidth, bb=28 28 566 566]{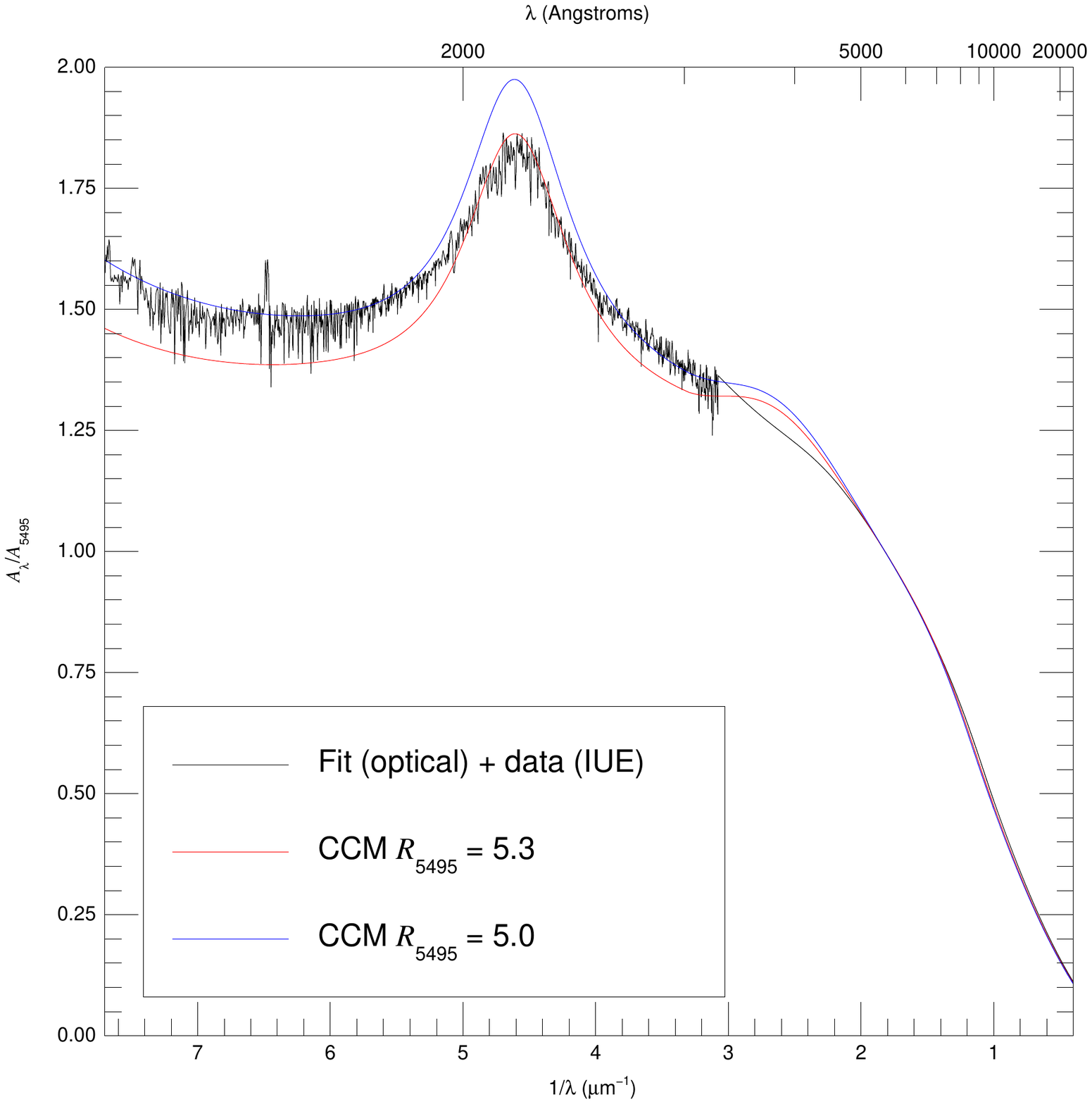}}
\caption{The Herschel 36 extinction law. The optical+NIR part is a CHORIZOS fit to the WFPC2 data using the new family of extinction laws.
The UV part is derived from the IUE data and the TLUSTY intrinsic SED obtained with CHORIZOS. Two CCM extinction laws are shown for comparison.}
\label{fig5}
\end{figure}

\section{Herschel 36 and its reflection nebula}

$\,\!$\indent Herschel 36 is the only bright point source in the FUV (Figure~\ref{fig2}), 
but surprisingly it contains only 29$\pm$6\% of the F170W flux in the PC.
Furthermore, there is only partial correlation between nebular emission and diffuse F170W (indicating that the cause is not the red leak). The latter comes 
  preferentially from the more extinguished regions: holes around Herschel 36 and NW part of the Hourglass.
The preferred explanation is that the FUV diffuse radiation is mostly (forward) scattered light (with smaller contributions from other sources such as
  free-free, free-bound, and 2-photon Lyman $\alpha$ continuum).
Our proposed geometry is shown in Figure~\ref{fig3}:

\begin{itemize}
 \item Herschel 36 is creating a cavity inside the molecular cloud.
 \item The direct light from the star arrives at us through a partially open hole.
 \item The bright regions of the Hourglass are farther away than Herschel 36 and they are the visible surface of the molecular cloud directly 
       illuminated by the star.
 \item The cloud is porous e.g. SW tunnel.
\end{itemize}

\section{Herschel 36: optical+NIR analysis}

$\,\!$\indent  We used CHORIZOS (\href{http://adsabs.harvard.edu/abs/2004PASP..116..859M}{Ma{\'\i}z Apell\'aniz 2004}, 
       \href{http://adsabs.harvard.edu/abs/2013hsa7.conf..657M}{2013b}) to measure the Herschel 36 extinction.
We fixed the luminosity class (5.0) and $T_{\rm eff}$ (38\,400 K). We left the amount ($E(4405-5495)$) and type ($R_{5495}$) of extinction and 
       logarithmic distance ($\log d$) as variables. 
We used two alternative extinction law families: CCM (\href{http://adsabs.harvard.edu/abs/1989ApJ...345..245C}{Cardelli et al. 1989}) and new 
       (\href{http://adsabs.harvard.edu/abs/2013hsa7.conf..583M}{Ma{\'\i}z Apell\'aniz 2013a}, 
       \href{http://adsabs.harvard.edu/abs/2014A&A...564A..63M}{Ma{\'\i}z Apell\'aniz et al. 2014}).
For the fit we used the WFPC2 F336W + F439W +  F487N + F502N + F547M + F656N + F673N + F675W + F814W and the 2MASS $J$ + $H$ filters. 
       The $K_s$-band photometry was excluded due to the IR excess caused by Herschel 36 C (see talk by J. Ma{\'\i}z Apell\'aniz).
The results are shown in Table~\ref{tab1} and Figure~\ref{fig4}): the new extinction laws provide a better fit to the optical+NIR data, especially for F336W.
It should be pointed out that the original CCM paper used Herschel 36 as an anchor point for large-$R_{5495}$ extinction laws but they overestimated the 
       amount of extinction ($E(B-V)$ = 0.89) and underestimated $R_{5495}$ (5.30).

\section{Herschel 36: onto the UV}

$\,\!$\indent We compared the IUE spectroscopy with the F170W + F255W magnitudes (Figure~\ref{fig4}). There is agreement between the IUE and WFPC2 fluxes
       a sign that the IUE extraction corresponds to the point source (it does not include the reflection nebula).
There is a discontinuity between the CCM fit to the optical+NIR data and the IUE spectroscopy at the UV-optical boundary. On the other hand, there is an
       agreement with the large-$R_{5495}$ extinction laws measured in 30 Doradus 
       (\href{http://adsabs.harvard.edu/abs/2014A&A...564A..63M}{Ma{\'\i}z Apell\'aniz et al. 2014}).
We calculated the UV extinction law $A_\lambda/A_{5495}$ by dividing the measured IUE flux by the intrinsic TLUSTY SED derived from CHORIZOS
       and the new extinction laws (Figure~\ref{fig5})> We found that:

\begin{itemize}
  \item The $R_{5495}$ = 5.0 CCM law works better than the $R_{5495}$ = 5.3 CCM law (the value used by CCM).
  \item The 2175~\AA\ bump is weaker than in the CCM laws.
  \item The most likely explanation for the above is that CCM used the wrong value of $E(4405-5495)$.
 \end{itemize}

\begin{table}
\caption{The multiple system Herschel 36. The separations, positions angles, and magnitude differences are all with respect to A.}
\centerline{$\,\!$}
\centerline{
\begin{tabular}{llcr@{.}lr@{.}l}
\hline
Component                & Other name        & Separation & \multicolumn{2}{c}{PA}             & \multicolumn{2}{c}{$\Delta$F814W} \\
                         &                   & (\arcsec)  & \multicolumn{2}{c}{(\degr)}        & \multicolumn{2}{c}{(mag)}         \\
\hline
Ba                       & KS 1-S            & 2.913      &   8&7                              & 6&98                              \\
Bb                       & KS 1-N            & 3.484      &   4&3                              & \multicolumn{2}{l}{$\sim$14}      \\
C$^*$                    & Herschel 36 SE    & 0.250      & 110&0                              & \multicolumn{2}{c}{---}           \\
D                        & G5.97-1.17        & 2.903      & 123&3                              & 6&40                              \\
E                        &                   & 0.740      & 201&7                              & 4&82                              \\
F                        &                   & 4.196      & 125&8                              & 10&0                              \\
G                        &                   & 2.774      & 272&6                              & 10&3                              \\
H                        &                   & 2.415      & 112&0                              & 10&4                              \\
I                        &                   & 3.456      & 168&2                              & 12&3                              \\
J                        &                   & 1.660      & 313&0                              & \multicolumn{2}{l}{$\sim$13}      \\
\hline
\end{tabular}
}

$^*$ Data from \href{http://adsabs.harvard.edu/abs/2006ApJ...649..299G}{Goto et al. (2006)}.
\label{tab2}
\end{table}

\section{Other sources in the Hourglass}

$\,\!$\indent There is only a handful of point sources seen in most optical filters in the PC.
We have measured the seporations, position angles, and $\Delta$F814W for the Herschel 36 multiple system (Table~\ref{tab2}), defined as the objects within 
     5\arcsec\ of A:
We have not attempted to resolve the embedded star Herschel 36 C (the contrast in the optical is very large and we are using aperture photometry), so the
     information provided in that case is from  \href{http://adsabs.harvard.edu/abs/2006ApJ...649..299G}{Goto et al. (2006)}.
We note that Herschel 36 Ba is much brighter than Bb in F814W (\href{http://adsabs.harvard.edu/abs/2006MNRAS.366..739A}{Arias et al. 2006}).

\begin{figure}
\centerline{\includegraphics*[width=1.00\linewidth]{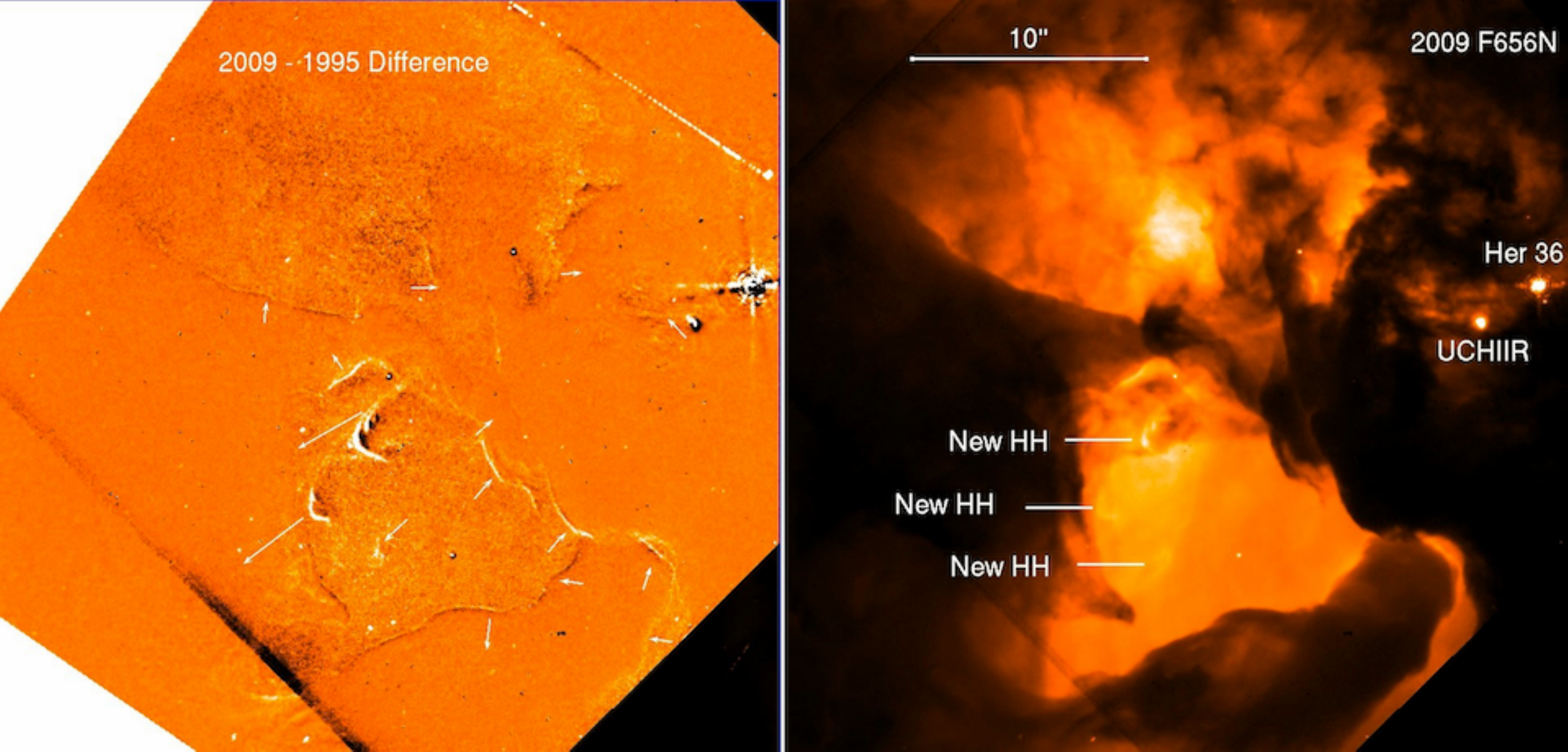}}
\caption{(left) Differential F656N image between 2009
and 1995. Whiter colours indicate stronger emission in
2009. The white arrows mark the direction of the detected movement. Their
lengths are proportional to the modulus of the motion vectors. North is up and east is left.
(right) F656N image of the same region. Herschel 36,
the ultracompact H\,{\sc ii} region G5.97-1.17, and the newly discovered Herbig-Haro objects are labelled.}
\label{fig6}
\end{figure}

Probably, the most interesting componente of Herschel 36 after A is D ($\equiv$ G5.97-1.17), 
an ultracompact H\,{\sc ii} region proposed as a proplyd by \href{http://adsabs.harvard.edu/abs/1998AJ....115..767S}{Stecklum et al. (1998)}.
For Herschel 36 D we measured a shift of $\sim$0.3 PC px (0\farcs015 or 19 AU) between 1995 and 2009 (Figure~\ref{fig6}): 
         this might correspond to the orbital motion around Herschel 36 with a minimum period of 17\,000 a. 
We also determined the extinction from the H$\alpha$/H$\beta$ ratio (F487N and F656N have low continuum contamination) assuming the same 
         $R_{5495}$ as for Herschel~36~A and obtained $A_{5495}$ = 9.9$\pm$0.3 mag.

\section{Herbig-Haro objects and nebular expansion}

$\,\!$\indent The comparison between new and archival WFPC2 images also reveals internal movements in the Hourglass Nebula. 
Three new Herbig-Haro objects are identified  in Figure~\ref{fig6}:
the HH nebular structures are displaced 0\farcs2-0\farcs3 during the interval between observations, indicating tangential velocities of 
       85-130 km/s. 
Also, a variety of nebular structures show displacements of up to 1 px (0\farcs045), suggesting an anisotropic expansion of the whole 
       Hourglass Nebula (Figure~\ref{fig6}).

\section*{References}
\begin{itemize}
 \item \href{http://adsabs.harvard.edu/abs/2006MNRAS.366..739A}{Arias, J. I. et al. 2006, {\it MNRAS} {\bf 366}, 739}.
 \item \href{http://adsabs.harvard.edu/abs/2010ApJ...710L..30A}{Arias, J. I. et al. 2010, {\it ApJL} {\bf 710}, 30}.
 \item \href{http://adsabs.harvard.edu/abs/1989ApJ...345..245C}{Cardelli, J., Clayton, G. C., \& Mathis, J. S. 1989, {\it ApJL} {\bf 345}, 245}.
 \item \href{http://adsabs.harvard.edu/abs/1994ApJ...435L..63G}{Gilliland, R. L. 1994, {\it ApJL} {\bf 435}, 63}.
 \item \href{http://adsabs.harvard.edu/abs/2006ApJ...649..299G}{Goto, M. et al. 2006, {\it ApJ} {\bf 649}, 299}.
 \item \href{http://adsabs.harvard.edu/abs/2003hstc.conf..346M}{Ma{\'{\i}}z Apell{\'a}niz, J. 2003, {\it 2002 HST Calibration Workshop}, 346}.
 \item \href{http://adsabs.harvard.edu/abs/2004PASP..116..859M}{Ma{\'{\i}}z Apell{\'a}niz, J. 2004, {\it PASP} {\bf 116}, 859}.
 \item \href{http://adsabs.harvard.edu/abs/2013hsa7.conf..583M}{Ma{\'{\i}}z Apell{\'a}niz, J. 2013a, {\it Highlights of Spanish Astrophysics VII}, 583}.
 \item \href{http://adsabs.harvard.edu/abs/2013hsa7.conf..657M}{Ma{\'{\i}}z Apell{\'a}niz, J. 2013b, {\it Highlights of Spanish Astrophysics VII}, 657}.
 \item \href{http://adsabs.harvard.edu/abs/2014A&A...564A..63M}{Ma{\'{\i}}z Apell{\'a}niz, J. et al. 2014, {\it A\&A} {\bf 564}, 63}.
 \item \href{http://adsabs.harvard.edu/abs/2014ApJS..211...10S}{Sota, A. et al. 2014, {\it ApJS} {\bf 211}, 10}.
 \item \href{http://adsabs.harvard.edu/abs/1998AJ....115..767S}{Stecklum, B. et al. 1998, {\it AJ} {\bf 115}, 767}.
\end{itemize}

\end{document}

%% file: mlaeff.tex
\def\hoy{\number\day \space de \space\ifcase\month\or
 Enero\or Febrero\or Marzo\or Abril\or Mayo\or Junio\or
 Julio\or Agosto\or Septiembre\or Octubre\or Noviembre\or Diciembre\fi
 \space de \number\year}
\def\ii/{\'{\i}}
\def\cion/{ci\'on}
\def\cao/{\c c\~ao}
\def\degr{\hbox{$^\circ$}}

\def\arcsec{\hbox{$^{\prime\prime}$}}
\def\utw{\smash{\rlap{\lower5pt\hbox{$\sim$}}}}
\def\udtw{\smash{\rlap{\lower6pt\hbox{$\approx$}}}}

\def\farcs{\hbox{$.\!\!^{\prime\prime}$}}

\def\tens#1{\ifmmode\mathchoice{\mbox{$\sf\displaystyle#1$}}
{\mbox{$\sf\textstyle#1$}}
{\mbox{$\sf\scriptstyle#1$}}
{\mbox{$\sf\scriptscriptstyle#1$}}\else
\hbox{$\sf\textstyle#1$}\fi}
\def\vec#1{\ifmmode\mathchoice{\mbox{\boldmath$\displaystyle#1$}}
{\mbox{\boldmath$\textstyle#1$}}
{\mbox{\boldmath$\scriptstyle#1$}}
{\mbox{\boldmath$\scriptscriptstyle#1$}}\else
\hbox{\boldmath$\textstyle#1$}\fi}
\def\bbbc{{\mathchoice {\setbox0=\hbox{$\displaystyle\rm C$}\hbox{\hbox
to0pt{\kern0.4\wd0\vrule height0.9\ht0\hss}\box0}}
{\setbox0=\hbox{$\textstyle\rm C$}\hbox{\hbox
to0pt{\kern0.4\wd0\vrule height0.9\ht0\hss}\box0}}
{\setbox0=\hbox{$\scriptstyle\rm C$}\hbox{\hbox
to0pt{\kern0.4\wd0\vrule height0.9\ht0\hss}\box0}}
{\setbox0=\hbox{$\scriptscriptstyle\rm C$}\hbox{\hbox
to0pt{\kern0.4\wd0\vrule height0.9\ht0\hss}\box0}}}}
\def\bbbq{{\mathchoice {\setbox0=\hbox{$\displaystyle\rm
Q$}\hbox{\raise
0.15\ht0\hbox to0pt{\kern0.4\wd0\vrule height0.8\ht0\hss}\box0}}
{\setbox0=\hbox{$\textstyle\rm Q$}\hbox{\raise
0.15\ht0\hbox to0pt{\kern0.4\wd0\vrule height0.8\ht0\hss}\box0}}
{\setbox0=\hbox{$\scriptstyle\rm Q$}\hbox{\raise
0.15\ht0\hbox to0pt{\kern0.4\wd0\vrule height0.7\ht0\hss}\box0}}
{\setbox0=\hbox{$\scriptscriptstyle\rm Q$}\hbox{\raise
0.15\ht0\hbox to0pt{\kern0.4\wd0\vrule height0.7\ht0\hss}\box0}}}}
\def\bbbt{{\mathchoice {\setbox0=\hbox{$\displaystyle\rm
T$}\hbox{\hbox to0pt{\kern0.3\wd0\vrule height0.9\ht0\hss}\box0}}
{\setbox0=\hbox{$\textstyle\rm T$}\hbox{\hbox
to0pt{\kern0.3\wd0\vrule height0.9\ht0\hss}\box0}}
{\setbox0=\hbox{$\scriptstyle\rm T$}\hbox{\hbox
to0pt{\kern0.3\wd0\vrule height0.9\ht0\hss}\box0}}
{\setbox0=\hbox{$\scriptscriptstyle\rm T$}\hbox{\hbox
to0pt{\kern0.3\wd0\vrule height0.9\ht0\hss}\box0}}}}
\def\bbbs{{\mathchoice
{\setbox0=\hbox{$\displaystyle     \rm S$}\hbox{\raise0.5\ht0\hbox
to0pt{\kern0.35\wd0\vrule height0.45\ht0\hss}\hbox
to0pt{\kern0.55\wd0\vrule height0.5\ht0\hss}\box0}}
{\setbox0=\hbox{$\textstyle        \rm S$}\hbox{\raise0.5\ht0\hbox
to0pt{\kern0.35\wd0\vrule height0.45\ht0\hss}\hbox
to0pt{\kern0.55\wd0\vrule height0.5\ht0\hss}\box0}}
{\setbox0=\hbox{$\scriptstyle      \rm S$}\hbox{\raise0.5\ht0\hbox
to0pt{\kern0.35\wd0\vrule height0.45\ht0\hss}\raise0.05\ht0\hbox
to0pt{\kern0.5\wd0\vrule height0.45\ht0\hss}\box0}}
{\setbox0=\hbox{$\scriptscriptstyle\rm S$}\hbox{\raise0.5\ht0\hbox
to0pt{\kern0.4\wd0\vrule height0.45\ht0\hss}\raise0.05\ht0\hbox
to0pt{\kern0.55\wd0\vrule height0.45\ht0\hss}\box0}}}}
\def\bbbz{{\mathchoice {\hbox{$\sf\textstyle Z\kern-0.4em Z$}}
{\hbox{$\sf\textstyle Z\kern-0.4em Z$}}
{\hbox{$\sf\scriptstyle Z\kern-0.3em Z$}}
{\hbox{$\sf\scriptscriptstyle Z\kern-0.2em Z$}}}}
\def\diameter{{\ifmmode\mathchoice
{\ooalign{\hfil\hbox{$\displaystyle/$}\hfil\crcr
{\hbox{$\displaystyle\mathchar"20D$}}}}
{\ooalign{\hfil\hbox{$\textstyle/$}\hfil\crcr
{\hbox{$\textstyle\mathchar"20D$}}}}
{\ooalign{\hfil\hbox{$\scriptstyle/$}\hfil\crcr
{\hbox{$\scriptstyle\mathchar"20D$}}}}
{\ooalign{\hfil\hbox{$\scriptscriptstyle/$}\hfil\crcr
{\hbox{$\scriptscriptstyle\mathchar"20D$}}}}
\else{\ooalign{\hfil/\hfil\crcr\mathhexbox20D}}%
\fi}}
\def\sq{\ifmmode\squareforqed\else{\unskip\nobreak\hfil
\penalty50\hskip1em\null\nobreak\hfil\squareforqed
\parfillskip=0pt\finalhyphendemerits=0\endgraf}\fi}
\def\squareforqed{\hbox{\rlap{$\sqcap$}$\sqcup$}}